\documentclass[11pt]{article} 
\usepackage[utf8]{inputenc}

\usepackage{amsfonts}
\usepackage{amsmath}
\usepackage{amssymb}
\usepackage{amsthm}
\usepackage{caption}
\usepackage{color}
    \definecolor{darkgreen}{rgb}{0,0.5,0}
    \definecolor{darkblue}{rgb}{0,0,0.6}
    \definecolor{purple}{rgb}{0.4,.2,0.7}
\usepackage[margin = 2.5cm]{geometry}
\usepackage{graphicx}
\usepackage[hyperfootnotes = false, colorlinks = true, linkcolor = darkblue, citecolor = blue,urlcolor  = darkblue]{hyperref}
\usepackage{subcaption}
\usepackage{appendix}
\usepackage{braket}
\usepackage{tikz}
\usepackage{mathrsfs} 
	\usepackage{booktabs}
\allowdisplaybreaks[4]

\begin{document}
\begin{titlepage}
	\renewcommand{\thefootnote}{\fnsymbol{footnote}}
\thispagestyle{empty}
\begin{center}
      ~\vspace{30mm}
    
    {\Large \bf {
   Signal of phase transition hidden in quasinormal modes of regular AdS black holes\\
} }

    \vspace{0.7in}
    
     {\bf Yang Guo${}^{}$\footnote{guoy@mail.nankai.edu.cn}, Hao Xie${}^{}$\footnote{xieh@mail.nankai.edu.cn}, and Yan-Gang Miao${}^{}$\footnote{Corresponding author: miaoyg@nankai.edu.cn}}

   \vspace{0.2in}

   {\em School of Physics, Nankai University, Tianjin 300071, China}
     
    \vspace{0.5in}

\end{center}

\vspace{0.5in}

\begin{abstract}

\end{abstract}

We discuss the intrinsic relations between thermodynamic phase transitions and quasinormal modes in regular AdS black holes, specifically in the Bardeen and Hayward AdS classes. To this end, we calculate the quasinormal modes of massless scalar field perturbations around small and large black holes via the Horowitz-Hubeny method. By investigating the isobaric and isothermal phase transitions for Bardeen and Hayward AdS black holes in detail, we observe that a dramatic change of quasinormal modes appears near the phase transition point of small and large black holes,  and that it corresponds to the swallow tail structure in the plane of Gibbs free energy with respect to pressure. Moreover, by analyzing the evolution of black holes along the coexistence curve of small and large black hole phases, we also observe  the dramatic change in quasinormal modes. Such a phenomenon confirms the signal of the phase transition in the quasinormal mode spectrum, which can be understood as a thermodynamic signal hidden in the dynamical spectrum.

\vspace{1in}


\setcounter{tocdepth}{3}


\end{titlepage}
\tableofcontents
\renewcommand{\thefootnote}{\arabic{footnote}}
\setcounter{footnote}{0}
\section{Introduction} 

 Quasinormal modes have been studied~\cite{Berti:2003ud,Konoplya:2019xmn,Guo:2020nci,Guo:2022hjp,Berti:2009kk,Konoplya:2011qq}  in a wide range of issues in  General Relativity and alternative theories of gravity. In general, quasinormal modes are used to describe the stability of black holes perturbed by an external field or by a spacetime metric, and they are also thought to carry  the information of gravitational waves.

There have been some arguments~\cite{Berti:2008xu,Liu:2014gvf} about whether black holes carry thermodynamic information that might be observed in future. In addition, it has been suggested~\cite{Jing:2008an} that scalar field perturbations are related to thermodynamic phase transitions of charged black holes, where an obvious change of quasinormal modes may be associated with a second order phase transition.
Furthermore, some studies have pointed out~\cite{Berti:2008xu,Rao:2007zzb} that a phase transition would probably connect to quasinormal modes, but their correspondence between the thermodynamic and dynamic aspects should be analyzed in more details instead of simply by numerical coincidence. 

Recently, a remarkable progress has been made~\cite{Wang:2000gsa,Shen:2007xk,Myung:2008ze,Koutsoumbas:2008pw,Mahapatra:2016dae}, indicating that the information of phase transitions does appear in the quasinormal mode spectra only for some specific cases in singular black holes.
It has been expected for a long time that thermodynamic phase transitions of black holes have some observational characteristics, which would be possible if the  phase transitions have the connections to quasinormal modes. In the present work, we deal with this issue for regular black holes through the procedure of phase transitions. In terms of the analyses~\cite{Kubiznak:2012wp,Xu:2023vyj,Wei:2015iwa,Wei:2014qwa} of thermodynamic phase transitions for AdS black holes, we use the semi-analytical method proposed~\cite{Horowitz:1999jd} by Horowitz and Hubeny 
to compute the quasinormal modes for two regular AdS black holes, the Bardeen and Hayward AdS classes, and then investigate the corresponding relations between  thermodynamic phase transitions  and quasinormal modes. In the thermodynamic aspect, we analyze the isobaric and isothermal phase transitions of small-large black holes in an extended phase space by showing the swallow tail structure in the plane of Gibbs free energy with respect to pressure, and also discuss the evolution of black holes along the coexistence curve of small and large black hole phases. On the other hand, in the dynamic aspect, we calculate the quasinormal modes of massless scalar field perturbations around small and large Bardeen and Hayward AdS black holes via the Horowitz-Hubeny method. By connecting the two aspects, we indeed find the signal of the phase transition in the quasinormal mode spectrum for Bardeen and Hayward AdS black holes.

Our paper is organized as follows. In Sec.~\ref{sec:pt1}, we make a general introduction to Einstein gravity coupled with nonlinear electrodynamics, and then analyze  thermodynamic phase transitions in regular AdS black holes in Sec.~\ref{sec:ptBH}.  Next, we introduce the massless scalar field perturbation in an asymptotic AdS spacetime in terms of the Horowitz-Hubeny method in Sec.~\ref{sec:ptHH}. We compute the quasinormal modes of scalar field perturbations when the Bardeen and Hayward AdS black holes undergo the isobaric and isothermal phase transitions in Sec.~\ref{subs:isobar} and Sec.~\ref{subc:isotherm}, respectively. In Sec.~\ref{sec:QNMcc}, we calculate quasinormal modes when the two black holes evolve along the coexistence curves, and analyze the behavior of quasinormal modes near the phase transition point of small-large black holes.  Finally, we give our conclusions in Sec.~\ref{sec:con}.

\section{Phase transition in regular AdS black holes}
\label{sec:pt}
\subsection{Einstein gravity coupled with  nonlinear electrodynamics}\label{sec:pt1}
 The Einstein gravity theory coupled with a nonlinear electromagnetic field is described by the action,
 \begin{eqnarray}
	S=\frac{1}{16\pi}\int d^4x\sqrt{- g}\left( { R}-2\Lambda-\mathcal{L}(F) \right),
\end{eqnarray}
	 where $R$ is the scalar curvature, $\Lambda$ the cosmological constant, and $F=\frac{1}{4} F^{ab} F_{ab}$ the electromagnetic invariant in the four-dimensional  spacetime with  indices $a$ and $b$. And the Lagrangian density of nonlinear electrodynamics is given~\cite{Fan:2016hvf} by
	\begin{eqnarray}
	\mathcal{L}(F)=\frac{4 \mu  M \left(2 F q^2\right)^{(\nu +3)/4}}{q^3 \left[ 1+\left(2F q^2\right)^{\nu /4}\right]^{(\mu +\nu)/\nu }},\label{eq:L}
\end{eqnarray}
which yields the static and spherically symmetric solution with the following shape function, 
\begin{eqnarray}
	f(r)=1-\frac{2Mr^{\mu-1}}{(r^\nu+q^{\nu})^{\mu/\nu}} +\frac{r^2}{l^2},\label{shapefunc}
\end{eqnarray}
where $M$ denotes the black hole mass, $q$ the magnetic charge, $l$ the radius of AdS spacetime, and $\mu$ and $\nu$ the dimensionless constants which characterize~\cite{Miskovic:2010ui,Miskovic:2010ey} the nonlinear degree of electromagnetic
fields. 
For the case of $\mu=3$ and $\nu=2$, this solution reduces~\cite{Ayon-Beato:2000mjt} to the Bardeen AdS black hole, and for the case of $\mu=\nu=3$, it reduces~\cite{Hayward:2005gi} to the Hayward AdS black hole. 
  The first law of thermodynamics can be written~\cite{Rasheed:1997ns,Zhang:2016ilt,Guo:2023pob} in the form,
  \begin{eqnarray}
  	d M=\frac{\kappa}{8\pi} d A+\Psi d q +  K_q d q +K_M d M,\label{1stlaw}
  \end{eqnarray}
where  
$\Psi$ is the magnetic potential, and $K_q$ and $K_M$ appear as the factors of the extra terms $K_q d q$ and $K_M d M$,
 \begin{eqnarray}
 	\Psi&=&\frac{M  \left\{\mu  q^{\nu } r_+^{-\nu }+3 \left( 1+ q^{\nu } r_+^{-\nu }\right) \left[\left(1+q^{\nu } r_+^{-\nu }\right)^{\mu /\nu }-1\right]\right\}}{2 q \left(1+q^{\nu } r_+^{-\nu }\right)^{(\mu +\nu)/\nu }},\label{Psi}\\ 
 	K_q&=& \frac{M r_+^{\mu } \left[3-3 \left(q^{\nu } r_+^{-\nu }+1\right)^{\mu /\nu }+\mu  q^{\nu }(r_+^{\nu }+ q^{\nu })^{-1}\right]}{2 q \left(r_+^{\nu }+ q^{\nu }\right)^{\mu /\nu }},\label{Kq} \\
 	K_M&=&1-\frac{r_+^{\mu }}{\left(r_+^{\nu }+q^{\nu }\right)^{\mu /\nu }}.\label{KM}
 \end{eqnarray}
Here $r_+$ stands for the radius of outer horizons as usual.  
 
\subsection{The small-large black hole phase transition}\label{sec:ptBH}
\subsubsection{Bardeen AdS black holes}\label{subsec:Bard}

The first law, see Eq.~\eqref{1stlaw} together with the term $VdP$ owing to the introduction of the AdS spacetime, can be rearranged to be a ``standard form", i.e. the formulation without the extra terms mentioned above, as follows:
\begin{eqnarray}
	d M=T' d S+ V' d P +\Psi' d q , \label{Bard1stLaw}
\end{eqnarray}
where the effective temperature $T'$, effective volume $V'$ and effective magnetic potential $\Psi'$ are defined by
\begin{eqnarray}
	T' \equiv \frac{T}{1-K_M},\qquad V'\equiv\frac{V}{1-K_M},\qquad \Psi' \equiv \frac{\Psi+K_q}{1-K_M}.\label{tpvpto}
\end{eqnarray}
For Bardeen AdS black holes with $\mu=3$ and $\nu=2$, by using Eqs.~(\ref{Psi}), (\ref{Kq}), (\ref{KM}), and (\ref{tpvpto}) we obtain
\begin{eqnarray}
	T' &=&\frac{\sqrt{q^2+r_+^2} \left(8 \pi  P r_+^4-2 q^2+r_+^2\right)}{4 \pi  r_+^4},\label{eq:Tprime}\\
	V'&=&\frac{4}{3} \pi  \left(q^2+r_+^2\right)^{3/2},\label{eq:Vprime}\\
	\Psi'&=&\frac{3 M q}{q^2+r_+^2}.\label{eq:BarT}
\end{eqnarray}
In the extended phase space, $(T', P, \Psi'; S, V', q)$, the Gibbs free energy takes the form,
\begin{eqnarray}
	G' = M-T' S, \label{eq:G}
\end{eqnarray}
and correspondingly the equation of state reads from Eqs.~(\ref{eq:Tprime}) and (\ref{eq:Vprime}),
\begin{eqnarray}
	P=	P(T', V')=\frac{4 \pi  T' \left(\frac{3 V'}{4 \pi }-q^3\right)^{4/3}+3 q^3-\frac{3 V'}{4 \pi }}{8 \pi  \left(\frac{3 V'}{4 \pi }-q^3\right)^{5/3}}.\label{eq:BarEOS}
\end{eqnarray}
Substituting Eqs.~(\ref{eq:Tprime}), (\ref{eq:Vprime}), and \eqref{eq:BarEOS} into the following critical condition,
\begin{eqnarray}
	\left( \frac{\partial P}{\partial V'}\right)_{T'}=0, \qquad \left( \frac{\partial^2 P}{\partial V'^2}\right)_{T'}=0,\label{eq:critical}
\end{eqnarray}
we can solve the critical point,
\begin{eqnarray}
	T'_c&=&\frac{5 \sqrt{188 \sqrt{10}-505}}{432 \pi  q},\nonumber\\
	V'_c&=&	\frac{4}{3} \left(2 \sqrt{10}+5\right)^{3/2} \pi  q^3,\nonumber\\
	P_c&=&\frac{5 \sqrt{10}-13}{432 \pi  q^2},
\end{eqnarray}
at which the phase transition of small and large black holes goes to the end in the evolution of the Bardeen AdS class, see Sec.~\ref{sec:QNMcc}, for the details.

\begin{figure}[ht]
	\centering		
	\includegraphics[width=.5\linewidth]{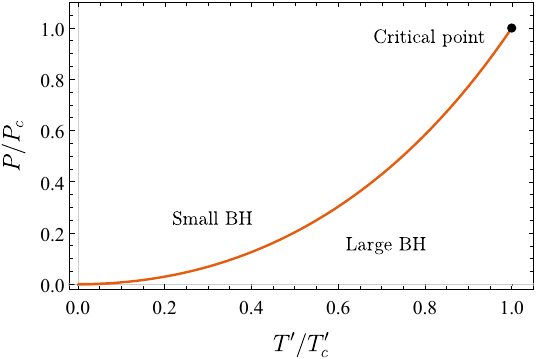}  
	\caption{The coexistence curve of small-large phases for Bardeen AdS black holes}
	\label{fig:barcoe}
\end{figure}

Fig.~\ref{fig:barcoe} clearly shows the existence of phase transitions of small and large  black holes  in the Bardeen AdS class, where the  coexistence curve given by the Gibbs free energy Eq.~(\ref{eq:G}) obeys strictly the Maxwell law in the extended phase space, $(T', P, \Psi'; S, V', q)$. The curve describes the coexistence of small and large black holes when the phase transition occurs. The critical point is highlighted by black color at the end of the coexistence curve. From this figure, we can see that the effective temperature and pressure of the small black hole phase are strictly the same as those of the large black hole phase.

\subsubsection{Hayward AdS black holes}
As discussed for the Bardeen AdS class, we derive the following effective quantities by using Eqs.~\eqref{1stlaw}-(\ref{tpvpto}) with $\mu=\nu=3$ for the Hayward AdS class, 
\begin{eqnarray}
	T' &=& \frac{8 \pi  P r_+^5-2 q^3+r_+^3}{4 \pi  r_+^4},\nonumber\\
	V'&=&\frac{4}{3} \pi  \left(q^3+r_+^3\right),\nonumber\\
	\Psi'& =& \frac{3Mq^2r_+^3}{(r_+^3+q^3)^2}.
\end{eqnarray}
Corresponding, we obtain the critical point of phase transitions,
\begin{eqnarray}
	T'_c&=&\frac{3}{8\sqrt[3]{20} \pi  q},\nonumber \\
	V'_c&=&28 \pi  q^3, \nonumber\\
	P_c&=&\frac{3}{80 \sqrt[3]{50}  \pi  q^2}.
\end{eqnarray}

\begin{figure}[ht]
	\centering		
	\includegraphics[width=.5\linewidth]{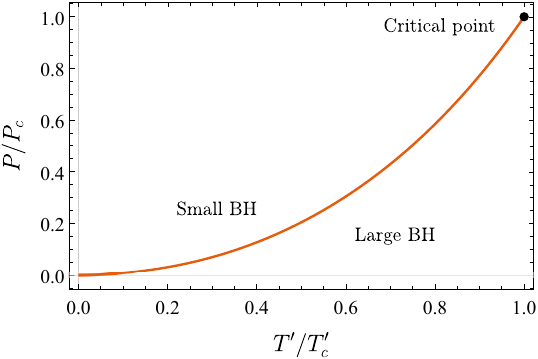}  
	\caption{The coexistence curve of small-large phases for Hayward AdS black holes}
	\label{fig:HC}
\end{figure}

Fig.~\ref{fig:HC} depicts the coexistence curve of small and large black hole phases in the Hayward AdS class, where the critical point is highlighted by black color at the end of the coexistence curve. Similarly, the effective temperature and pressure  of small black holes are the same as those of large black holes when the phase transition occurs.

 \section{Signal of phase transition hidden in quasinormal modes}
 \label{sec:ptQNM}

\subsection{Perturbation in an asymptotic AdS spacetime}
\label{sec:ptHH}

Now we introduce the Horowitz-Hubeny method~\cite{Horowitz:1999jd} for a scalar field perturbation in an asymptotic AdS spacetime. A massless scalar field is described by the equation, 
\begin{equation}
	{\nabla}^2 \Phi =0, \label{KG}
\end{equation}
and it is usually decomposed by the following form in a spherically symmetric spacetime,
\begin{equation}
	\Phi(t,r,\theta,\phi)= \frac{1}{r} \psi (r) Y (\theta, \phi) e^{-i \omega t},
\end{equation}
where $ Y (\theta, \phi)$
is spherical harmonics. Using the tortoise coordinate defined by $dr_{*}=dr/f(r)$, we derive the radial equation,
\begin{equation}
	\frac{d^2 \psi}{{dr_{*}}^2}+(\omega^2-V) \psi =0,
\end{equation}
where the potential takes the form for the metric with the shape function Eq.~(\ref{shapefunc}),
\begin{equation}
	V(r)=f(r)\left[ \frac{l(l+1)}{r^2}+\frac{1}{r} f'(r)\right] .
\end{equation}

In an asymptotically flat spacetime, the potential of scalar perturbations vanishes at infinity. Therefore, the quasinormal modes are determined by the outgoing waves near infinity, $\Phi \sim e^{-i \omega (t-r_{*})}$, and by the ingoing waves near an outer horizon, $\Phi \sim e^{-i \omega (t+r_{*})}$.
In an asymptotic AdS spacetime, however, the potential of scalar perturbations diverges~\cite{Horowitz:1999jd} at infinity and thus the wave function $\Phi$ vanishes. We note that the potential of gravitational perturbations tends~~\cite{Cardoso:2001bb,Chatzifotis:2021pak,Vlachos:2021weq}  to  a non-zero and non-infinite constant in an asymptotic AdS spacetime. As a result, the quasinormal modes of scalar perturbations are determined only by ingoing waves near an outer horizon in an AdS spacetime. To this end, we calculate the modes of ingoing waves near an outer horizon, $\Phi \sim e^{-i \omega (t+r_{*})}$, and write the line element in the ingoing Eddington coordinates for the sake of convenience,
\begin{equation}
	ds^2=-f(r)dv^2+2dv dr+r^2 d \Omega^2.
\end{equation}
By using the corresponding decomposition of  $\Phi$,
\begin{equation}
	\Phi (v, r, \theta, \phi )=\frac{1}{r} \psi (r) Y(\theta, \phi) e^{-i \omega v},\label{sep}
\end{equation}
and substituting it into Eq.~(\ref{KG}), we derive the following radial equation,
\begin{equation}
	f(r) \frac{d^2 \psi (r)}{dr^2}+\left[f'(r)-2i \omega \right] \frac{d\psi (r)}{dr}-V(r) \psi (r) =0,\label{radial}
\end{equation}
where the effective potential $V(r)$ is expressed by
\begin{equation}
	V(r)=\frac{1}{r} f'(r) + \frac{l(l+1)}{r^2}.\label{eq:poten}
\end{equation}

In order to calculate the quasinormal modes, we expand the solution in a power series at the horizon $r_+$ and restrict the solution to zero at infinity. The range of our solution is $r_+<r<+\infty$. In order to give the quasinormal modes numerically, we change the range to a finite one by setting $x=1/r$, and thus transfer Eq.~(\ref{radial}) into the following form,
\begin{equation}
	s(x) \frac{d^2 \psi (x)}{dx^2}+\frac{t(x)}{x-x_+} \frac{d \psi (x)}{dx}+\frac{u(x)}{(x-x_+)^2} \psi (x) =0,\label{x}
\end{equation}
where $x_+=1/r_+$ and the three coefficient functions take the forms,
\begin{align}
	s(x)&=-\frac{ x^4 f(x)}{x-x_+},\\
	t(x)&=-2x^3 f(x) - x^4 \frac{d f(x)}{dx} - 2i \omega x^2,\\
	u(x)&=(x-x_+) V(x).
\end{align}
We can  expand them as the Taylor series with the center at $x=x_+$, 
\begin{eqnarray}
s(x)=\sum_{n=0}^\infty s_n (x-x_+)^n, \nonumber\\
t(x)=\sum_{n=0}^\infty t_n (x-x_+)^n, \nonumber\\
u(x)=\sum_{n=0}^\infty u_n (x-x_+)^n.\label{coeexp}
\end{eqnarray}

It is worth noting that $s_0=2x_+^2 \kappa$, $t_0=2x_+^2 (\kappa -i \omega)$, and $u_0=0$ at the horizon $x=x_+$, where $\kappa$ is the surface gravity and can be expressed as $\kappa=f'(r_+)/2$. To determine the behavior of the  quasinormal modes near the horizon, we set the solution $\psi(x)$ as
\begin{equation}
	\psi(x)=\sum_{n=0}^{\infty} a_n (x-x_+)^n.\label{ps}
\end{equation}
When substituting Eq.~(\ref{coeexp}) and Eq.~(\ref{ps}) into Eq.~(\ref{x}), we obtain the recursion relations of $a_n$,
\begin{equation}
	a_n=-\frac{1}{P_n} \sum_{k=0}^{n-1} \left[k(k-1) s_{n-k} + k t_{n-k} + u_{n-k}\right]a_k,
\end{equation}
where $P_n$ is given by
\begin{equation}
	P_n=n(n-1) s_0 + n t_0 =2 x_+^2 n(n \kappa -i \omega).
\end{equation}
Since the solution $\psi(x)$  satisfies the boundary condition: $\psi(x)\rightarrow0$ in the limit of $x\rightarrow0$, we need to find the zeros of the equation, $\sum_{n=0}^\infty a_n(\omega) (-x_+)^n=0$, in the complex $\omega$ plane.  As a result, we compute the complex quasinormal frequencies by truncating the series, $\sum_{n=0}^\infty a_n(\omega) (-x_+)^n$, at a large enough value of $n$ and computing the relevant finite sum as a function of $\omega$. In the following subsections, we calculate the quasinormal frequencies to the order of $n=50$ and obtain the convergent numerical results, where our investigation follows the procedures of isobaric and isothermal phase transitions in Sec.~\ref{subs:isobar} and Sec.~\ref{subc:isotherm}, respectively, and also the evolution procedure along coexistence curves of small-large black hole phases in Sec.~\ref{sec:QNMcc}.

\subsection{Isobaric phase transition}\label{subs:isobar}
\begin{figure}[h]
	\centering
	\begin{subfigure}[b]{0.48\textwidth}
		\centering
		\includegraphics[width=\textwidth]{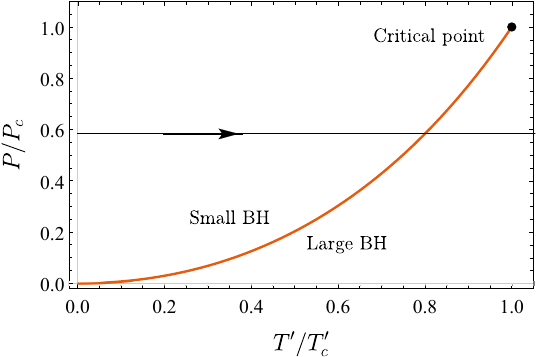}
		\caption{Bardeen AdS black hole}\label{sbfig:Biso}
	\end{subfigure}
	\begin{subfigure}[b]{0.482\textwidth}
		\centering
		\includegraphics[width=\textwidth]{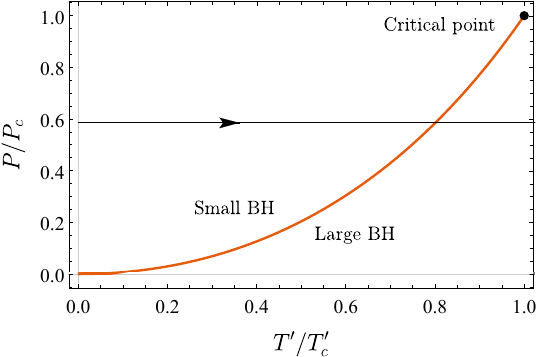}
		\caption{Hayward AdS black hole}\label{sbfig:Hiso}
	\end{subfigure}
	\caption{ Isobaric phase transition and coexistence curve for small and large black holes }
	\label{fig:BHCi}
\end{figure}

When the effective temperature goes large, the Bardeen and Hayward AdS black holes evolve and cross the coexistence curve along the isobar as shown in Fig.~\ref{fig:BHCi}, where the arrow indicates the direction of small-large black hole phase transition, that is, the small black hole phase passes through the phase transition point of the coexistence curve and then turns into the large black hole phase.
    
Next, we calculate the quasinormal modes when the Bardeen and Hayward AdS black holes evolve along this isobar shown in Fig.~\ref{fig:BHCi}, and list the numerical results of quasinormal mode frequencies for both the Bardeen and Hayward AdS classes in Table~\ref{tab:isobar}. For each class, the upper part, above the horizontal line, gives the data for the small black hole phase, while the lower part does for the large black hole phase. Specifically, the isobar $P=0.00483$ or $P=0.00757$ is set for Bardeen or Hayward AdS black holes. When the effective temperature of small black holes increases and reaches the value at which the small-large black hole phase transition occurs, $T'_0=0.05577$ for the Bardeen AdS class and $T'_0=0.07036$ for the Hayward AdS class, respectively, a dramatic change appears in quasinormal mode spectra.

\begin{table}[h]
	\centering
		\begin{tabular}{ p{3em} cc}
		\hline
		\hline

			\multicolumn{3}{c}
			{Bardeen AdS black hole}\\
			\hline
			$T'$ & $r_+$ & $\omega$  \\
			\hline
			0.050 & 0.915 & 2.35954 - 3.60483 i  \\
			0.051 & 0.927 & 2.37202 - 3.62750 i \\
			0.052 & 0.941 & 2.38627 - 3.65313 i \\
			0.053 & 0.956 & 2.40127 - 3.68049 i \\
			0.054 & 0.974 & 2.41947 - 3.71606 i \\
			0.055 & 0.994 & 2.43878 - 3.75648 i \\
			\bottomrule[1pt]
			0.056 & 3.442 & 6.47910 - 9.38757 i \\
			0.057 & 3.699 & 6.94789 - 10.0555 i \\
			0.058 & 3.920 & 7.35130 - 10.6318 i \\
			0.059 & 4.119 & 7.71520 - 11.1521 i \\
			0.060 & 4.304 & 8.05349 - 11.6366 i \\
			\hline
			\hline
			\multicolumn{3}{c}
			{Hayward AdS black hole}\\
			\hline
			$T'$ & $r_+$ & $\omega$  \\
			\hline
			  0.065 & 0.816 & 2.55852 - 2.91317 i \\
			0.066 & 0.825 & 2.56397 - 2.92913 i \\
			0.067 & 0.835 & 2.56989 - 2.94712 i \\
			0.068 & 0.846 & 2.57623 - 2.96714 i \\
			0.069 & 0.858 & 2.58288 - 2.98917 i \\
			0.070 & 0.871 & 2.58983 - 3.01316 i \\
			\bottomrule[1pt]
			0.071 & 2.888 & 5.68775 - 7.75504 i \\
			0.072 & 3.044 & 5.96104 - 8.16489 i \\
			0.073 & 3.183 & 6.20559 - 8.53076 i \\
			0.074 & 3.311 & 6.43156 - 8.86816 i \\
			0.075 & 3.431 & 6.64403 - 9.18482 i \\
		\hline
		\hline
		\end{tabular}
	\caption{The quasinormal mode frequencies of a massless scalar field perturbation with respect to the effective temperature $T'$, where the isobar $P=0.00483$ or $P=0.00757$ is set for Bardeen or Hayward AdS black holes. 
	The phase transition point is located at  $T'_0=0.05577$ for Bardeen AdS black holes, or $T'_0=0.07036$ for Hayward AdS black holes.}\label{tab:isobar}
\end{table}

\subsection{Isothermal phase transition}\label{subc:isotherm}
\begin{figure}[h]
 	\centering
 	\begin{subfigure}[b]{0.48\textwidth}
 		\centering
 		\includegraphics[width=\textwidth]{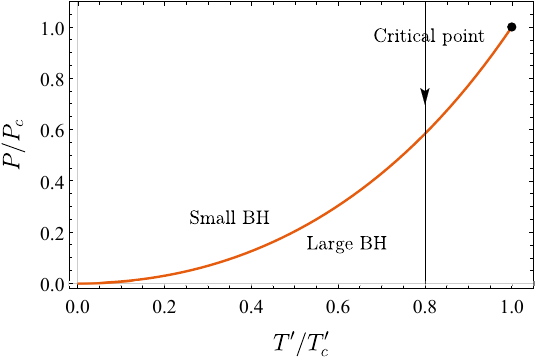}
 		\caption{Bardeen AdS black hole}\label{sbfig:BC}
 	\end{subfigure}
 	\begin{subfigure}[b]{0.482\textwidth}
 		\centering
 		\includegraphics[width=\textwidth]{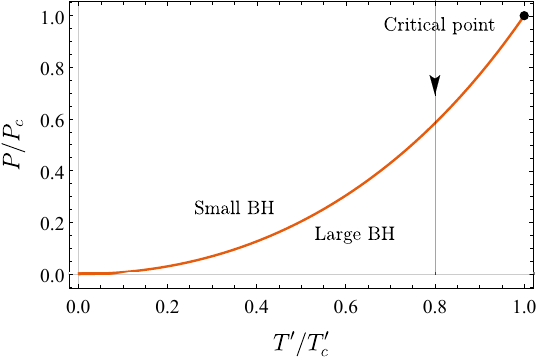}
 		\caption{Hayward AdS black hole}\label{sbfig:HC}
 	\end{subfigure}
 	\caption{Isothermal phase transition and coexistence curve for small and large black holes}
 	\label{fig:BHC}
 \end{figure}

When the pressure becomes small, the Bardeen and Hayward AdS black holes evolve and cross the coexistence curve along the isotherm as shown in Fig.~\ref{fig:BHC}, where the arrow indicates the direction of small-large black hole phase transition, that is, the small black hole phase passes through the phase transition point of the coexistence curve and then turns into the large black hole phase.

Next, we calculate the quasinormal modes when the Bardeen and Hayward AdS black holes evolve along this isotherm shown in Fig.~\ref{fig:BHC}, and list the numerical results of quasinormal mode frequencies for both the Bardeen and Hayward AdS classes in Table~\ref{tab:isotherm}. For each class, the upper part, above the horizontal line, gives the data for the small black hole phase, while the lower part does for the large black hole phase. Specifically, the isotherm $T'=0.8T'_c$ is set for both Bardeen and Hayward AdS black holes. When the pressure of small black holes decreases and reaches the value at which the small-large black hole phase transition occurs, $P_0=0.00483$ for the Bardeen AdS class and $P_0=0.00757$ for the Hayward AdS class, respectively, a dramatic change appears in quasinormal mode spectra.
 
Combining the results in the above subsection with those in this one, we can conclude that the same phenomenon exists, i.e., the dramatic change of quasinormal modes appears in both the isobaric and isothermal phase transitions.

\begin{table}[h]
 	\centering
 	\begin{tabular}{ p{3em} cc}
 		\hline
 		\hline
 		\multicolumn{3}{c}{Bardeen AdS black hole}\\
 		\hline
 		$P$ & $r_+$ & $\omega$  \\
 		\hline
 	    0.0053 & 0.988 & 2.43274 - 3.74402 i \\
 		0.0052 & 0.993 & 2.43791 - 3.75502 i \\
 		0.0051 & 0.997 & 2.44169 - 3.76321 i \\
 		0.0050 & 1.002 & 2.44628 - 3.77294 i \\
 		0.0049 & 1.008 & 2.45140 - 3.78524 i \\
 		\bottomrule[1pt]
 		0.0048 & 3.443 & 6.48103 - 9.39022 i \\
 		0.0047 & 3.639 & 6.83841 - 9.89935 i \\
 		0.0046 & 3.829 & 7.18522 - 10.3943 i \\
 		0.0045 & 4.015 & 7.52490 - 10.8800 i \\
 		0.0044 & 4.202 & 7.86690 - 11.3693 i \\
 		0.0043 & 4.390 & 8.21091 - 11.8621 i \\
 		\hline
 		\hline
 		\multicolumn{3}{c}{Hayward AdS black hole}\\
 		\hline
 		$P$ & $r_+$ & $\omega$  \\
 		\hline
 		0.0080 & 0.866 & 2.58719 - 3.00396 i \\
 		0.0079 & 0.868 & 2.58825 - 3.00760 i \\
 		0.0078 & 0.871 & 2.58983 - 3.01316 i \\
 		0.0077 &0.873  & 2.59087 - 3.01684 i \\
 		0.0076 & 0.876 & 2.59240 - 3.02232 i \\
 			\bottomrule[1pt]
 		0.0075 & 2.848 & 5.61789 - 7.65011 i \\
 		0.0074 & 2.948 & 5.79271 - 7.91257 i \\
 		0.0073 & 3.045 & 5.96280 - 8.16752 i \\
 		0.0072 & 3.141 & 6.13160 - 8.42015 i \\
 		0.0071 & 3.236 & 6.29907 - 8.67041 i \\
 		0.0070 & 3.331 & 6.46693 - 8.92091 i \\
 		\hline
 		\hline
 	\end{tabular}
 	\caption{The quasinormal mode frequencies of a massless scalar field perturbation with respect to the pressure $P$, where the isotherm $T'=0.8T'_c$ is set for both Bardeen and Hayward AdS black holes. 
 	The phase transition point is located at  $P_0=0.00483$ for Bardeen AdS black holes, or $P_0=0.00757$ for Hayward AdS black holes.}\label{tab:isotherm}
 \end{table}

\subsection{Quasinormal modes on coexistence curves} \label{sec:QNMcc}

\begin{figure}[h]
	\centering
	\begin{subfigure}[b]{0.48\textwidth}
		\centering
		\includegraphics[width=\textwidth]{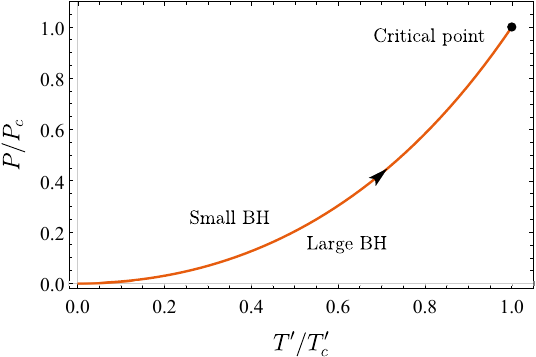}
		\caption{Bardeen AdS black hole}\label{sbfig:BA}
	\end{subfigure}
	\begin{subfigure}[b]{0.482\textwidth}
		\centering
		\includegraphics[width=\textwidth]{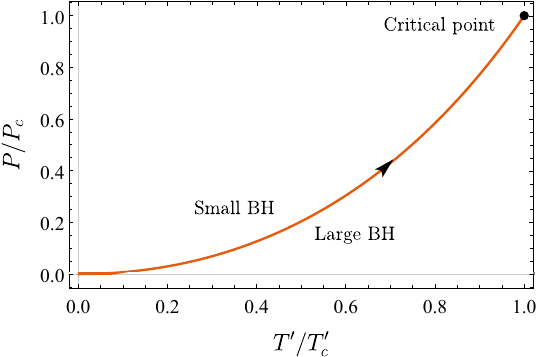}
		\caption{Hayward AdS black hole}\label{sbfig:HA}
	\end{subfigure}
	\caption{Coexistence curve for small and large black holes}
	\label{fig:BHA}
\end{figure}

In Sec.~\ref{subs:isobar} and Sec.~\ref{subc:isotherm}, we investigate the isobaric and isothermal phase transitions for both Bardeen and Hayward AdS black holes and find that the dramatic changes of quasinormal modes indeed indicate the signal of phase transitions. In this subsection, we analyze the thermodynamic and dynamic behaviors of the two regular AdS black holes when the evolution of the black holes goes along the coexistence curves as shown in Fig.~\ref{fig:BHA}, where the arrow shows the evolution direction of the black holes towards the critical points.

\begin{figure}[h]
	\centering
	\begin{subfigure}[b]{0.477\textwidth}
		\centering
		\includegraphics[width=\textwidth]{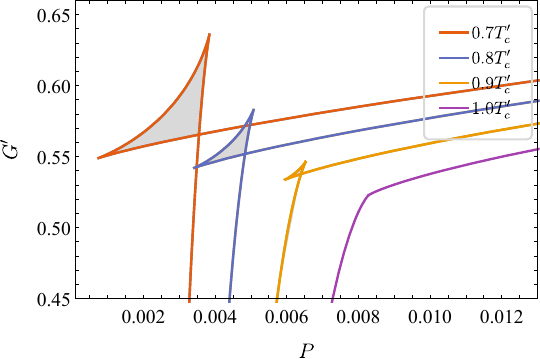}
		\caption{Bardeen AdS black hole}\label{sbfig:BarG}
	\end{subfigure}
	\begin{subfigure}[b]{0.493\textwidth}
		\centering
		\includegraphics[width=\textwidth]{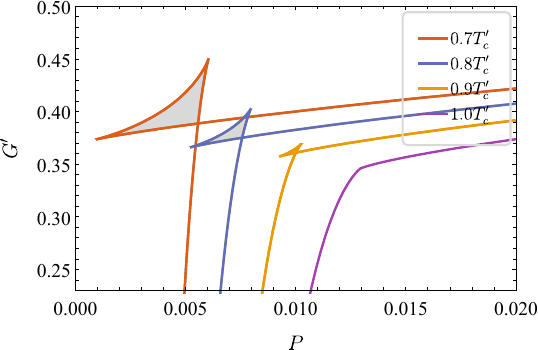}
		\caption{Hayward AdS black hole}\label{sbfig:HayG}
	\end{subfigure}
	\caption{The swallow tail structure of Gibbs free energy}
	\label{fig:BHG}
\end{figure}

As known, when a phase transition occurs in an AdS black hole,  the Gibbs free energy of this black hole exhibits the swallow tail structure as shown in Fig.~\ref{fig:BHG}, which corresponds to an intermediate state of phase transitions. In our case of Bardeen and Hayward AdS classes, this intermediate state means that the small and large black holes coexist. When the critical point is reached, see the purple curve in Fig.~\ref{fig:BHG}, the small-large black hole phase transition ends and then both the intermediate state and  the swallow tail structure of the Gibbs free energy  disappear simultaneously.

\begin{figure}[h]
	\centering
	\begin{subfigure}[b]{0.48\textwidth}
		\centering
		\includegraphics[width=\textwidth]{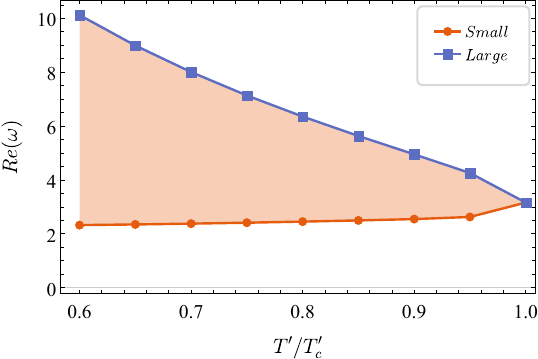}
	\end{subfigure}
	\begin{subfigure}[b]{0.493\textwidth}
		\centering
		\includegraphics[width=\textwidth]{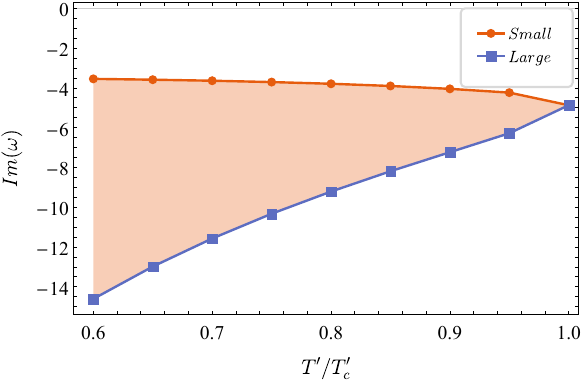}
	\end{subfigure}
	\caption{Quasinormal modes when the Bardeen AdS black hole evolves along the coexistence curve}
	\label{fig:BarRI}
\end{figure}

\begin{figure}[h]
  	\centering
  	\begin{subfigure}[b]{0.48\textwidth}
  		\centering
  		\includegraphics[width=\textwidth]{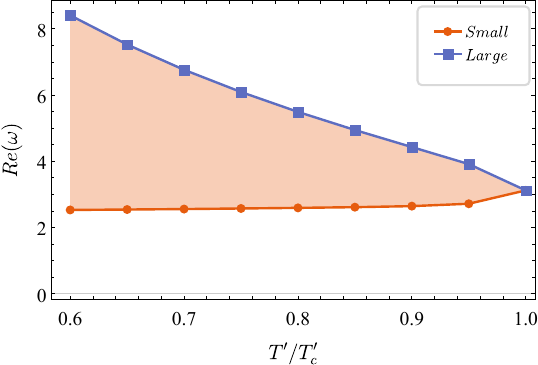}
  	\end{subfigure}
  	\begin{subfigure}[b]{0.496\textwidth}
  		\centering
  		\includegraphics[width=\textwidth]{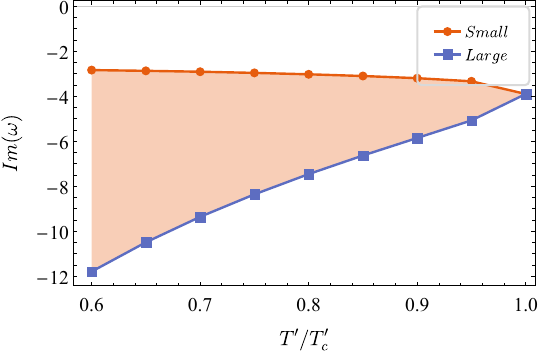}
  	\end{subfigure}
  	\caption{Quasinormal modes when the Hayward AdS black hole evolves along the coexistence curve}
  	\label{fig:HayRI}
  \end{figure}

By calculating those quasinormal modes that the Bardeen and Hayward AdS black holes evolve along the coexistence curves, we find from Fig.~\ref{fig:BarRI} and Fig.~\ref{fig:HayRI} that the differences between small and large black holes' real (imaginary) parts are very dramatic when the black holes stay at the point of coexistence curves far away from the critical point, and that such a difference gradually decreases when the point moves towards the critical point and vanishes at the critical point. Moreover, the shaded area in the swallow tail structure of Gibbs free energy, see Fig.~\ref{fig:BHG}, corresponds to the shaded area associated with quasinormal modes, see Fig.~\ref{fig:BarRI} and Fig.~\ref{fig:HayRI}, and both areas disappear at the critical point of phase transitions. As a result, the dynamic behavior of small-large black hole coexistence confirms the signal of thermodynamic  phase transitions from an alternative viewpoint that is different from isobaric and isothermal phase transitions.
 
\begin{figure}[h]
	\centering
	\begin{subfigure}[b]{0.496\textwidth}
		\centering
		\includegraphics[width=\textwidth]{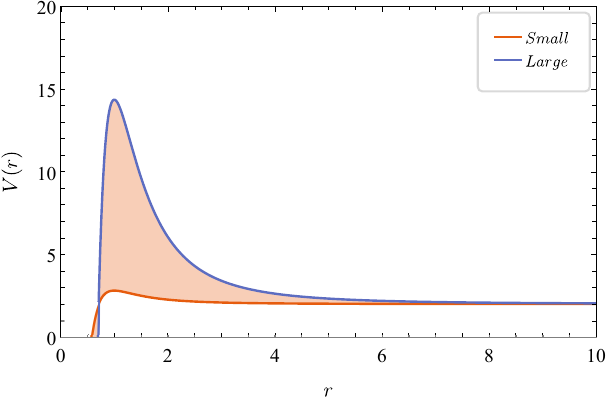}
			\caption{Bardeen AdS black hole}\label{sbfig:Vbar}
	\end{subfigure}
	\begin{subfigure}[b]{0.496\textwidth}
		\centering
		\includegraphics[width=\textwidth]{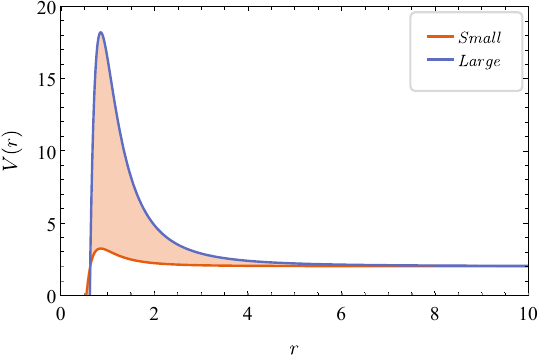}
		\caption{Hayward AdS black hole}\label{sbfig:Vhay}
	\end{subfigure}
	\caption{The potentials of small and large black holes}
	\label{fig:Poten}
\end{figure}

At the end of this section, we clarify the reason for the drastic changes in the quasinormal mode spectra of small and large black holes. We point out that such drastic changes originate from the sensitivity~\cite{Nollert:1996rf} of quasinormal modes to the changes of potentials rather than from spectral instabilities, see, for instance, some interesting discussions~\cite{Jaramillo:2020tuu,Cheung:2021bol,Destounis:2023ruj} on spectral instabilities. In order to affirm the former, we plot the effective potential Eq.~\eqref{eq:poten} in Fig.~\ref{fig:Poten}.
We can see that the effective potential in the Eddington coordinates is non-zero and non-infinite at infinity, indicating that the Horowitz-Hubeny method is valid~\cite{Horowitz:1999jd}, and that it changes dramatically between small and large black holes. When the phase transition of small and large black holes occurs, there is a large gap between the effective potential of small black holes and the effective potential of large black holes, which leads to the drastic changes of quasinormal mode spectra. On the other hand, in order to rule out the latter, we multiply Eq.~\eqref{radial} by $\bar{\psi}(r)$ (complex conjugate of ${\psi}(r)$) and integrate from $r_+$ to $\infty$~\cite{Horowitz:1999jd},
\begin{eqnarray}
	\int_{r_+}^{\infty}dr\left[f(r)|\psi'(r)|^2+V(r)|\psi(r)|^2\right]=-\frac{|\omega|^2 |\psi(r_+)|^2}{{\rm Im}(\omega)},
\end{eqnarray}
which clearly shows that there are no solutions to unstable modes with Im$(\omega) \geq0$.

   \section{Summary} 
   \label{sec:con}
   In the present work, we investigate systematically the correlation between phase transitions and quasinormal modes for regular AdS black holes by following the procedures of isobaric and isothermal phase transitions and the evolution procedure along coexistence curves of small-large black hole phases, where the phase transitions belong to thermodynamics, while the quasinormal modes dynamics. As expected, we
   confirm that such a correlation is inevitable but not accidental by numerical coincidence. 
   We thus conclude that quasinormal modes can be a dynamic probe of thermodynamic phase transitions, which implies that the information of phase transitions may be hidden in the observations of gravitational waves that are closely related to quasinormal modes.

 \paragraph{Acknowledgments}
 The authors would like to thank  the anonymous referee for the helpful comments that improve this work greatly.
 This work was supported in part by the National Natural Science Foundation of China under Grant No. 12175108. 
      

\bibliographystyle{JHEP}
\bibliography{references}

\end{document}